\documentclass[%
 reprint,
 amsmath,amssymb,
 aps,
]{revtex4-2}

\usepackage{graphicx}% Include figure files
\usepackage{epstopdf}
\usepackage{dcolumn}% Align table columns on decimal point
\usepackage{bm}% bold math
\usepackage{physics}
\usepackage{mathrsfs}
\usepackage{mathtools}
\usepackage{xcolor}
\usepackage{xfrac}
\usepackage{float}
\usepackage{hyperref}% add hypertext capabilities
\usepackage{booktabs}
\usepackage{multirow}
\usepackage{threeparttable}

\def\dd{\text{d}}
\def\pp{\partial}
\def\be{\begin{equation}}
\def\ee{\end{equation}}
\def \mq{\mathbf{q}}
\def \mp{\mathbf{p}}

\newcommand{\lb}{\left(}
\newcommand{\rb}{\right)}
\newcommand{\lsq}{\left[}
\newcommand{\rsq}{\right]}
\newcommand{\lc}{\left\{}
\newcommand{\rc}{\right\}}
\newcommand{\mf}{\mathbf}
\newcommand*\diff{\mathop{}\!\mathrm{d}}
\newcommand{\avg}[1]{\left\langle #1 \right\rangle}

\begin{document}

\preprint{APS/123-QED}

\title{The Use of Wigner Expansion and Path Integral Method for Quantum Correction of the Low-Order Spectral Moments and Collision-Induced Absorption Profiles} 
% Force line breaks with \\

\author{Daniil N. Chistikov}
\email{Corresponding author: danichist@yandex.ru}
\author{Artem A. Finenko}

\affiliation{A.M. Obukhov Institute of Atmospheric Physics, Russian Academy of Sciences, 3 Pyzhevsky per., Moscow 119017, Russia} 
\affiliation{Institute of Quantum Physics, Irkutsk National Research Technical University, 83 Lermontov Street, Irkutsk 664074, Russia}

\date{\today}

\begin{abstract}
This study aims at examination of the lower-order spectral moments in collision-induced absorption (CIA), taking the translational He-Ar band as an example. General quantum corrections for the zeroth and first spectral moments are derived on the basis of the Wigner expansion up to the order of $\hbar^8$ and $\hbar^{10}$, respectively. These corrections were then explored for numerical simulation of the He-Ar moments over the temperature range 50-500 K. The accuracy of the obtained temperature dependencies is validated through the comparison with direct quantum solution of the scattering and bound-state problems as well as with the results obtained using the path integral (PI) approach. Quite satisfactory agreement was achieved as a result of the application of the two independent ways of approximate accounting for the quantum nature of absorption. The robustness of the Wigner expansion and PI corrections was thus demonstrated, both of which can be used to obviate direct quantum simulation of the CIA spectral moments. The approximate method is suggested for an accurate simulation of the CIA spectral profile on the basis of the obtained quantum-corrected moments and the classical trajectory-based formalism.
\end{abstract}

\maketitle

%\tableofcontents
\section{Introduction}

% Collision induced absoption originates from... \\
% Molecular pairs \\
% Application to the planetary atmospheres \\
% Need for lower temperatures for several atmospheres\\
% Desymmetrization at lower temperatures fails\\
% Correction of the profile by moments sometimes works (ch4-co2 as an example).
% Moments are easier computed by quantum corrections and pimc than by quantum calculations

Weak interaction between distinguishable atoms in the gas phase is known to induce temporary variations in the pair's dipole moment. In case of rare gas mixtures this effect gives rise to the so-called translational collision-induced absorption bands in the far infrared, which are observable at elevated gas densities \cite{Bosomworth1965, Bukhtoyarova1977}. 
The importance of taking CIA into account for planetary and astrophysical investigations is presently well recognized \cite{Chubb2024, Abel2011}. The study of such integrated CIA characteristics as low-order spectral moments is known to be of great value for developing accurate theoretical methods, which would enable simulation of`reliable CIA spectral profiles for subsequent atmospheric or other applications.  

The most accurate description of CIA spectra is provided by the quantum-mechanical consideration.
At the same time, this approach is associated with high computational complexity, especially for systems with a large number of degrees of freedom. At present, the CIA spectra that were modeled using fully quantum-mechanical approach are reported only for the systems consisting from two interacting atoms or two diatomics (e.g., H$_2-$H$_2$ \cite{Abel2011} and N$_2-$N$_2$ \cite{Karman2015}). 
% At least for the diatomics, the use of huge computational resources is required for CIA simulation in the far infrared.
As a result, various approximations, such as e.g. reducing the complicated intermolecular potential to the isotropic one \cite{Abel2011, ElKader2014}, are largely used. Although the isotropic approximation substantially simplifies numerical calculations, it can lead to significant deviation of a result from the measured data \cite{Turbet2020}. The trajectory-based approach \cite{Chistikov2021,Finenko2022} and, to some extent, classical many-body perspectives \cite{Fakhardji2021, Hartmann2017} proved themselves as practical alternatives to quantum-mechanical consideration. However, both approaches relying on classical dynamics of molecules in a gas, have to employ correction of the obtained spectral profiles to make them satisfy the quantum principle of detailed balance. This correction referred to as desymmetrization procedure is inherently approximate, as there is no direct and general correspondence between the spectral profiles obtained using classical and quantum perspectives. Consequently, various desymmetrization procedures have been proposed, which,
although yielding similar estimates in the classical limit ($\sfrac{h\nu}{k_\text{B} T} \ll 1$), can lead to significantly different estimates in cases when quantum effects are pronounced ($\sfrac{h\nu}{k_\text{B} T} \gg 1$). For instance,  it was demonstrated in Ref. \cite{Finenko2022} that various desymmetrization procedures can result in significantly disparate intensity distribution of the CH$_4-$N$_2$ induced spectrum in the 200$-$500 cm$^{-1}$ range at temperatures below 100 K. This discrepancy is particularly noteworthy in the context of analyzing the radiance spectrum of Titan's atmosphere.  

The CIA band can be quantitatively characterized using the lower-order spectral moments, which describe the intensity distribution within the band \cite{Frommhold2006}. In the present study we focus our attention on the zeroth and first spectral moments, which are static properties amenable for quantum-statistical consideration using Wigner expansion in powers of $\hbar$ or path integral (PI) approach. The He-Ar prototype system was chosen by us for this study bearing in mind that quantum effects are expected to be substantial in a system with lightweight helium. The spectral moments are examined over the temperature range of 50$-$500 K that is typical for the most of planetary atmospheres.
Our suggested methods have not been systematically considered earlier in the literature in order to quantify the integrated CIA characteristics. Because of this, we confronted our estimates to those obtained with the result of solution of the time-independent quantum-mechanical problem. The practicality of the use of the Wigner expansion or PI approaches is evaluated, considering that these methods yield series of estimates in powers of $\hbar$ or necklace size, respectively. We believe that once a reliable procedure of the lower-order spectral moments calculation is developed, it can help introducing proper correction to the classically simulated spectral profile as well.

Quantum corrections are widely employed in the calculation of the virial coefficients, with convenient expressions for the first rotational and translational corrections having been derived by \citet{pack1983first} and \citet{wormer2005second}. These corrections have also been successfully applied to various static thermodynamic properties of molecular gases and fluids \cite{barocchi1985quantum, Barocchi1987}, diatomic molecules \cite{prudente2001calculation} and in plasma physics \cite{Gombert1998}. Furthermore, quantum corrections have been studied for the collision-induced light scattering \cite{barocchi1981quantum} and CIA \cite{hartye1975moment} spectral moments. Expressions for the zeroth, first and second spectral moments for the general spectrum up to $\hbar^4$ have been derived \cite{Barocchi1982}. Notably, \citet{Frommhold2006} demonstrated that corrections up to $\hbar^2$ for the spectral moments are sufficient to reproduce quantum-mechanical results for room-temperature rare gas mixtures.

%The moments obtained using phase-space integration should match those obtained from simulated spectra through following the evolution of the dipole moment either in an ensemble of pair trajectories or in large collection of molecules, as in molecular dynamics simulation. Keep in mind that the spectral profile's shape is altered when a desymmetrization is subsequently applied, which affects its spectral moments.

This paper is organized as follows. Section \ref{sec:theory} outlines the theoretical framework for modeling CIA profile in rare gas mixtures from both quantum and classical perspectives. The Wigner expansion and path integral approaches employed for calculation of the lower-order spectral moments are described in subsections \ref{subsec:wigner} and \ref{subsec:path-integral}, respectively. The results obtained employing quantum-mechanical, quantum-statistical, and classical approaches are then discussed and compared in Section \ref{sec:discussion}, where the advantages of each approach are highlighted. Finally, subsection \ref{subsec:trajectory} proposes a method that enables applying quantum corrections to the spectral profile calculated with classical trajectory-based formalism.

\section{Theory and computation}
\label{sec:theory}

\subsection{Collision-induced spectra}

Collision-induced binary absorption coefficient is given by the following expression \cite{Frommhold2006}:
\begin{equation}
\label{eq:alpha_to_J_relation}
    \frac{\alpha(\nu)}{\rho_1\rho_2} =  \frac{8\pi^3 N_L^2}{3\hbar}\nu \lsq 1 - \exp \lb -\frac{\hbar c \nu}{k_\text{B} T} \rb \rsq V J(\nu),
\end{equation}
where $\nu$ is the wavenumber, $N_L$ is the Loschmidt constant, $\hbar$ is the reduced Planck constant, $k_\text{B}$ is the Boltzmann constant and $J(\nu)$ is the spectral density.

Within time-independent quantum-mechanical formulation, the computation of the collision-induced spectrum is reduced to solving the Schr\"odinger equation, which for the diatomic molecular system takes the form:

\begin{gather}
\begin{aligned}
    & \lsq -\frac{\hbar^2}{2m}\frac{\dd^2}{\dd R^2} + \frac{\hbar^2\ell(\ell+1)}{2m R^2} + U(R)  \right] \psi = E \psi,
\end{aligned}
  \label{eq:radial-schroedinger}
\end{gather}
where $R$ is the interparticle distance, $m$ is the diatomic reduced mass $m = m_{A}m_{B}/(m_{A}+m_{B})$ for the pair AB, $\ell$ is the partial wave quantum number, and $U(R)$ is the interaction potential. % angular momentum quantum number. 
Here, the kinetic energy of center of mass is separated and only the relative motion of atoms is taken into consideration. 
The radial quantum states are characterized by the energy and the angular momentum quantum number: $\psi_i(R) \equiv \ket{E_i,\ell_i}$.

% Spectral density $J(\nu)$ can be divided into four components, associated with the transitions between bound states ($J_{bb}$), between free states ($J_{ff}$) and between bound and free states ($J_{bf}$ and $J_{fb}$). These components are calculated through quantum mechanical approach in the following way \cite{Frommhold2006}:
The spectral density $J(\nu)$ consists of four components: $J_\textrm{ff}$ (free-to-free transitions), $J_\textrm{bb}$ (bound-to-bound transitions), and $J_\textrm{bf}$ and $J_\textrm{fb}$ (bound-to-free and free-to-bound transitions). These components can be calculated according to \cite{Frommhold2006}:

\begin{widetext}
\begin{align}
  V J_\textrm{ff} &= \lambda_0^3 \hbar \sum_{\ell_i, \ell_f}(2\ell_i + 1) C(\ell_i 1 \ell_f;000)^2 
  \times \int\limits_0^{\infty} \exp \lb  -\frac{E_i}{k_\text{B} T} \rb \Big| \matrixel{E_i + \hbar\omega, \ell_f}{\mu}{E_i\ell_i} \Big|^2 \diff{E_i},  
  \label{eq:J_ff_quan} \\
  V J_\textrm{bb} &= \lambda_0^3 \hbar \sum_{\ell_i, \ell_f}(2\ell_i + 1) C(\ell_i 1 \ell_f; 000)^2 \times
  \sum_{v_i, v_f} \exp \lb  -\frac{E_{v_i \ell_i}}{k_\text{B} T} \rb \Big| \matrixel{E_{v_f \ell_f} \ell_f}{\mu}{E_{v_i, \ell_i} \ell_i} \Big|^2  \delta \lb E_{v_f\ell_f} - E_{v_i\ell_i} - \hbar\omega \rb,
  \label{eq:J_bb_quan} \\
  V J_\textrm{bf} &= \lambda_0^3 \hbar \sum_{\ell_i, \ell_f}(2\ell_i + 1) C(\ell_i 1 \ell_f;000)^2 \times \sum_{v_i} \exp \lb  -\frac{E_{v_i \ell_i}}{k_\text{B} T} \rb \Big| \matrixel{E_{v_i \ell_i} + \hbar \omega, \ell_f}{\mu}{E_{v_i, \ell_i} \ell_i} \Big|^2,
  \label{eq:J_bf_quan} \\
  V J_\textrm{fb} &= \lambda_0^3 \hbar \sum_{\ell_i, \ell_f}(2\ell_i + 1) C(\ell_i 1 \ell_f; 000)^2 \times \sum_{v_f} \exp \lb -\frac{E_{v_f \ell_f} - \hbar\omega}{k_\text{B} T} \rb \Big| \matrixel{E_{v_f \ell_f} \ell_f}{\mu}{E_{v_f \ell_f} - \hbar\omega, \ell_i} \Big|^2,
  \label{eq:J_fb_quan}
\end{align}
\end{widetext}
where $C(\ell_i 1 \ell_f;000)$ is the Clebsch-Gordan coefficient. The summation in  Eqs. (\ref{eq:J_bb_quan}-\ref{eq:J_fb_quan}) includes all bound states $v_i, v_f$. According to the selection rules, transition probabilities are non-zero only when the angular momentum quantum numbers differ by 1, i.e., $\ell_f = \ell_i \pm 1$. The term $\mu \equiv \mu(R)$ represents the radial part of the induced dipole moment of the diatomic system:
\be
  \mu_{\nu}(\mathbf{R}) = \sqrt{\frac{4\pi}{3}} Y_{1 \nu}(\hat{\mathbf{R}}) \mu(R), \quad \nu = 0, \pm 1.
\ee
In most of the cases the intensity of a collision-induced band for weakly interacting systems is dominated by ``free-free" transitions, the contribution of which gradually increases with temperature. However, in the lower temperatures domain or for the systems with relatively deep intermolecular potential well the contribution from true bound states can be substantial.

\subsection{Classical calculation}

The quantum-mechanical approach provides the most accurate description of induced absorption spectra, thereby allowing all the weak spectral effects to be accounted for. 
%At the same time, its full-strength application for the systems having several degrees of freedom requires extreme consumption of computational resources. As a result, the calculations are often constrained by the use of various approximations, of which the use of an isotropic assumption for the intermolecular potential \cite{abel2011collision, el2014spectral} is the most frequent. Although the isotropic approximation substantially simplifies numerical calculations, it can lead to discrepancies with the results of observations for molecular systems with notable anisotropy \cite{turbet2020measurements}. High computational requirements preclude presently the full quantum CIA spectra simulation for anistropically interacting pairs larger than two interacting linear molecules \cite{karman2015collision,karman2015quantum}.
An alternative way for CIA spectral simulation explores classical mechanical treatment of molecular pair dynamics  \cite{Buryak2014,Hartmann2017,Chistikov2019,Chistikov2021,Fakhardji2021}. In a series of publications, it has been shown that the trajectory-based approach provides an efficient and reliable tool for CIA spectral simulation in various systems, from rare gas mixtures \cite{Buryak2014} to more complicated N$_2$-N$_2$ \cite{Serov2024,Finenko2024}, CO$_2$-CO$_2$ \cite{Odintsova2021}, and CH$_4$-N$_2$ \cite{Finenko2022}.

In order to obtain the working equations for the classical approach, let us represent the spectral density function as a Fourier transform of the dipole autocorrelation function instead of the sum over matrix elements in Eqs. (\ref{eq:J_ff_quan}-\ref{eq:J_fb_quan}) \cite{McQuarrie}:

\be
  J(\nu) = \frac{1}{2\pi}\int\limits_{-\infty}^{\infty}C(t)e^{-2\pi i c \nu t} \dd t,
\ee
where $C(t)$ is the quantum dipole autocorrelation function:
\iffalse
\be
C(t) = \int\mathllap{\sum_i} e^{ -E_i/kT}\matrixel{i}{\boldsymbol{\mu}(0)\cdot e^{i\mathscr{H}t/\hbar}\boldsymbol{\mu}(0) e^{-i\mathscr{H}t/\hbar}}{i}
\ee
\fi
\be
  C(t) = \sum_i e^{-E_i/kT} \matrixel{i}{\boldsymbol{\mu}(0)\cdot e^{i\mathscr{H}t/\hbar}\boldsymbol{\mu}(0) e^{-i\mathscr{H}t/\hbar}}{i}.
  \label{eq:quantum_corr_func}
\ee
Here, $\mathscr{H}$ is the Hamiltonian of the system. The summation in Eq. \eqref{eq:quantum_corr_func} is valid over true bound states, whereas for free states, it should be reformulated as an integral.

Within classical formalism, the dipole autocorrelation function is expressed as follows:
\be 
  C_\textrm{cl}(t) = \frac{V}{Q}\int \boldsymbol{\mu}(0)\cdot\boldsymbol{\mu}(t)e^{-H/kT} \dd \Gamma,
  \label{eq:classical_corr_func}
\ee
where $Q$ denotes the classical partition function of the molecular system, $V$ is the volume of gas, and $H$ is the classical Hamiltonian. The integration is carried out over the phase space of the molecular system, with $\dd\Gamma$ representing the volume element in the phase space. The details on the trajectory-based CIA simulation can be found in Refs. \cite{Chistikov2019,Chistikov2021}. In evaluating Eq. (\ref{eq:classical_corr_func}), the integration is conducted via Monte Carlo sampling, while the time evolution of the dipole moment is calculated by propagating Hamilton equations of motion for the selected molecular system. 

The primary difference between the spectral profiles derived from the quantum and classical approaches is that they satisfy distinct detailed balance conditions \cite{Frommhold2006}. The quantum mechanical detailed balance originates from the different thermal population of the initial and final states:
\be
  \label{eq:quantum_detbal}
  J_{\mathrm q}(-\nu) = \exp \lb -\frac{h c \nu}{k_\text{B} T} \rb J_{\mathrm q}(\nu).
\ee
In contrast, classical spectral function $J_\mathrm{cl}(\nu)$ has to be symmetric in the frequency domain due to reversibility of classical trajectories:
\be
  \label{eq:classical_detbal}
  J_{\mathrm{cl}}(-\nu) =  J_{\mathrm{cl}}(\nu).
\ee
This difference means that the classical spectral function must be subjected to a desymmetrization procedure prior to being compared with experimental data \cite{Frommhold2006}. Once desymmetrized, the spectral function satisfies the quantum detailed balance condition. Various desymmetrization procedures have been proposed. For the far infrared spectral range, the procedures of Sch\"ofield \cite{Schofield1960} 
\be
  J_\text{D3}(\nu) = J_\mathrm{cl}(\nu) \exp \lb \frac{h c \nu}{2 k_\text{B} T} \rb
  \label{eq:des-schofield}
\ee
and Frommhold \cite{Frommhold2006}
\begin{gather}
\begin{aligned}
  J_\text{D4a}(\nu) = &\frac{C_\text{cl}(0)}{C_\text{cl}(\beta h / \sqrt{2})}  \exp \lb \frac{h c \nu}{2 k_\text{B} T} \rb \times \\ 
  &\int\limits_{-\infty}^{+\infty} C_\text{cl} \lb \sqrt{t^2 + \lb \beta \hbar / 2 \rb^2} \rb e^{-2 \pi i c \nu t} \diff{t} 
\end{aligned}
  \label{eq:des-frommhold}
\end{gather}
were shown to result in a satisfactory agreement with both experimental data and the profiles obtained using quantum mechanical formalism \cite{Chistikov2019,Chistikov2021,Finenko2022}. The latter procedure \eqref{eq:des-frommhold} stems from an extension of the Egelstaff's procedure \cite{Egelstaff1962}.

\subsection{Spectral moments}

In order to characterize intensity distribution in the spectral profile, the so-called spectral moments are often used \cite{Chistikov2021,Frommhold2006}. The $n$-th spectral moment $\mathcal{M}_n$ is defined as an integral over the spectral density weighted with the $n$-th power of the frequency:
\be
  \mathcal{M}_n = \frac{(2\pi)^3N_L^2}{3\hbar}\int\limits_{-\infty}^{\infty} \nu^n J(\nu) \diff{\nu}.
\ee
Spectral moments can be expressed as integrals from the absorption coefficient provided spectral density is replaced by $\alpha(\nu)$ using Eq. \eqref{eq:alpha_to_J_relation} and the corresponding quantum \eqref{eq:quantum_detbal} or classical \eqref{eq:classical_detbal}  detailed balance conditions are taken into account. As a result we have
\be
  \label{eq:mom_q_alpha}
  \mathcal{M}_n^{\mathrm{q}} = \begin{dcases}
  \int\limits_{0}^{\infty}\nu^{n-1}\coth\left(\frac{hc\nu}{2kT}  \right)\alpha(\nu)\dd \nu,\\
  \int\limits_{0}^{\infty}\nu^{n-1}\alpha(\nu)\dd \nu \\
  \end{dcases}
\ee
for the quantum detailed balance and
\be
\label{eq:mom_cl_alpha}
 \mathcal{M}_n^{\mathrm{cl}} = \begin{dcases}
 2\int\limits_{0}^{\infty}\nu^{n-1}\left[ 1- \exp\left(-\frac{hc\nu}{kT}  \right)\right]\alpha(\nu)\dd \nu,\\
 0.\\
 \end{dcases}
\ee
for the classical one.
The expressions in Eqs. \eqref{eq:mom_q_alpha} and \eqref{eq:mom_cl_alpha} are given in a form, with the first line applying to even values of $n$ and the the second line applying to odd values. It is noteworthy that in the classical limit all odd spectral moments vanish, reflecting the classical spectral function symmetry in the frequency domain. 

Equivalent representation of the spectral moments can be also obtained in terms of the quantum averages of the operators involving dipole moment as \cite{Frommhold2006}
% The general expression for the $n$-th spectral moment can be written in two equivalent forms \cite{Frommhold2006}:
\be
\label{eq:Mn_commutator}
  \mathcal{M}_n^\text{q} = V\frac{8\pi^3N_L^2}{3(2\pi\hbar c)^{n+1}}\frac{1}{4\pi\varepsilon_0}\mathrm{Tr}\big\{  \hat{\rho} \boldsymbol{\mu}\cdot \smash[b]{ \underbrace{[\mathscr{H},\ldots[\mathscr{H},}_\text{$n$ times}\:\boldsymbol{\mu}]]}\big\}
\ee
or
\be
  \mathcal{M}_n^\text{q} =   V\frac{8\pi^3N_L^2}{3\hbar}i(2\pi i c)^{-n-1} \frac{1}{4\pi \epsilon_0}\left\langle\boldsymbol{\mu}(0)\cdot\frac{d^n}{dt^n}\boldsymbol{\mu}(t)  \right\rangle\bigg|_{t=0},
\ee
where 
\be
  \label{eq:rho_canonical}
  \hat{\rho} = \exp(-\beta\hat{\mathscr{H}})/\mathrm{Tr}\left[\exp(-\beta\hat{\mathscr{H}})\right] 
\ee
is the canonical density matrix. % In the above equation $\beta = 1/k_BT$.
Specifically, the zeroth and first spectral moments can be calculated using the following relations:
\begin{widetext}
\begin{align}
   \mathcal{M}_0^\text{q} &= \frac{4 \pi^2 N_L^2}{3\hbar c}\frac{1}{4\pi\varepsilon_0}4\pi\int\limits_0^\infty(\mu(R))^2g^{(0)}(R)R^2 \:\dd R, \label{eq:zeroth_g0}\\
  \mathcal{M}_1^\text{q} &= \frac{4\pi^2 N_L^2}{3m c^2}\frac{1}{4\pi\varepsilon_0}\int\limits_0^\infty\left\{\left(\frac{\dd \mu}{\dd R}\right)^2+\frac{2}{R^2}(\mu(R))^2\right\}g^{(0)}(R)R^2 \:\dd R,\label{eq:first_g0}
\end{align}
\end{widetext}
where $g^{(0)}(R)$ is the pair distribution function:
\be
\label{eq:g0_quantum}
  g^{(0)}(R)=\lambda_0^3\int\limits_0^\infty\dd E\sum_{\ell}\frac{2\ell+1}{4\pi}e^{-E/kT}\frac{1}{R^2}|\psi(R;E,\ell)|^2.
\ee
In the classical limit, the first spectral moment vanishes and the zeroth spectral moment can be calculated using Eq. \eqref{eq:zeroth_g0} with the pair distribution function given by:
\be
  g^{(0)}_{\mathrm{cl}}(R) = \exp\left( -\frac{U(R)}{k_BT} \right).
\ee

Spectral moments are valuable quantities that can be readily computed using both quantum and classical statistical methods. However, for molecular systems with multiple degrees of freedom, fully quantum mechanical calculations often become prohibitively expensive. This motivates the exploration of alternative approaches to CIA simulation, which could potentially offer a similar level of accuracy to purely quantum mechanical methods but at a significantly lower computational cost. The Wigner transform and the Feynman path integral are two prominent statistical approaches that have been developed to improve the accuracy of classical approximation. In the following sections, we explore the application of these methods to the retrieval of CIA spectral moments, using a relatively simple diatomic system as a case study.

\subsection{Wigner expansion}
\label{subsec:wigner}

In classical statistical mechanics, observables are characterized by phase-space distributions of coordinates and momenta, which evolve according to the classical equations of motion. In contrast, in quantum mechanics one deals with the wave functions and probabilities, expressed in either the coordinate or momentum representation. A direct comparison between these two approaches is hindered by the non-commutativity of quantum position and momentum operators, which prevents the existence of a joint probability distribution for coordinates and momenta. 

The quasi-probability function proposed by E. Wigner \cite{wigner1932quantum, hillery1984distribution, case2008wigner} provides a bridge between classical and quantum description of observables. In the Wigner formalism, the mean value of an operator $\hat{A}$ for a system in a pure state, described by a density matrix $\hat{\rho} = \ket{\psi}\bra{\psi}$, can expressed in terms of the Wigner function $P_w(\mf{q}, \mf{p})$:
\be
\label{eq:A_mean_value}
\mathrm{Tr}(\hat{\rho}\hat{A}) = \int\dd \mathbf{q}\int \dd \mathbf{p} A_w(\mathbf{q},\mathbf{p})P_w(\mathbf{q},\mathbf{p}),
\ee
where
\be
\label{eq:A_wigner}
A_w(\mathbf{q},\mathbf{p}) = \int\dd \mathbf{z} \, e^{i\mathbf{p}\mathbf{z}/\hbar} \matrixel{\mathbf{q}-\frac{1}{2}\mathbf{z}}{\hat{A}}{\mathbf{q}+\frac{1}{2}\mathbf{z}}
\ee
and
\be 
\label{eq:Pw_wigner}
P_w(\mathbf{q},\mathbf{p}) = \left(\frac{1}{\pi\hbar}\right)^s\int\limits_{-\infty}^{+\infty} \dd \mathbf{y} \, \psi^{\ast}(\mathbf{q}+\mathbf{y})\psi(\mathbf{q}-\mathbf{y}) e^{2i\mathbf{p}\mathbf{y}/\hbar},
\ee
where $s$ denotes the number of degrees of freedom.
When integrated over coordinates or momenta, the Wigner function represents the correct quantum mechanical probability distribution in momentum or coordinate space, respectively. At first glance, Eq. (\ref{eq:A_mean_value}) may suggest that the Wigner function $P_w$ is a direct quantum analogue of a classical phase space distribution function. In reality, this is not entirely accurate, as the function $P_w$ can take on negative values for a given $\mf{q}$ and $\mf{p}$, whereas classical distribution function values are always non-negative.

In the Wigner formalism, the mean value of the product of two general operators $\hat{A}$ and $\hat{B}$ can be formulated similarly to Eq. \eqref{eq:A_mean_value}:
\be
  \label{eq:Wigner_AB_trace}
  \mathrm{Tr}(\hat{A}\hat{B}) = \frac{1}{(2\pi \hbar)^s}\int \dd \mathbf{q}\int\text{d}\mathbf{p} \, A_w(\mathbf{q},\mathbf{p}) B_w(\mathbf{q},\mathbf{p}),
\ee
where $A(\mathbf{q},\mathbf{p})$ and $B(\mathbf{q},\mathbf{p})$ are the Wigner transforms of the operators $\hat{A}$ and $\hat{B}$, respectively, which can be obtained as prescribed by Eq. \eqref{eq:A_wigner}.

As demonstrated by Wigner in his seminal paper \cite{wigner1932quantum}, the reformulation of quantum mechanics in terms of the Wigner function provides a clear framework for constructing  quantum corrections to thermodynamic averages.
To apply this framework, we can substitute the canonical density matrix, given by Eq. \eqref{eq:rho_canonical}, as the operator $\hat{B}$ in Eq. (\ref{eq:Wigner_AB_trace}). Consequently,  corresponding Wigner equivalent, $\rho_w$, is expanded into a power series in $\hbar$. 
A convenient approach to constructing these approximations involves the Bloch equation for the operator $\hat{\Omega} = \exp(-\beta\hat{\mathscr{H}})$, which is defined by the following system of equations:
\be
  \begin{dcases}
  \frac{\pp \hat{\Omega}}{\pp \beta} = -\hat{\Omega}\hat{\mathscr{H}}, \\
  \hat{\Omega}(\beta = 0) = \hat{I}.
  \end{dcases}
\ee
The Wigner equivalent of this equation is given by
\be
  \label{eq:Omega_w_cos}
  \frac{\partial\Omega_w}{\partial\beta} = -H_w(q,p)\cos(\hbar\Lambda/2)\Omega_w(q,p),
\ee
where the operator $\Lambda$ is defined as
\be
  {\Lambda} = \overleftarrow{\frac{\pp}{\pp \mathbf{p}}}\overrightarrow{\frac{\pp}{\pp \mathbf{q}}} - \overleftarrow{\frac{\pp}{\pp \mathbf{q}}}\overrightarrow{\frac{\pp }{\pp \mathbf{p}}}.
\ee
In this expression, left (right) arrows indicate that the corresponding derivative acts on the function to the left (right) of the $\Lambda$ operator. The equation \eqref{eq:Omega_w_cos} is derived using the following relation established by \citet{GROENEWOLD1946405}, which relates the product of operators to its Wigner function counterpart
\be
\label{eq:wigner_multiplication}
\hat{F} = \hat{A}\hat{B} \quad \Leftrightarrow \quad F(q,p) = A(q,p) \exp \lb \frac{\hbar}{2i}\Lambda \rb B(q,p).
\ee

By expanding the cosine and $\Omega_w$ in Eq. (\ref{eq:Omega_w_cos}) in a power series in $\hbar$ and equating the terms with corresponding powers of $\hbar$, we obtain a set of differential equations. Solving these equations yields the sought-after terms $\chi_2$, $\chi_4$, etc., which appear in the power series expansion of the unnormalized density matrix \cite{imre1967wigner}:
\be
\label{eq:Omega_expansion}
  \Omega_w(\mathbf{q},\mathbf{p}) = (1 + \hbar^2 \chi_2 + \hbar^4\chi_4 + \ldots)\exp(-\beta H),
\ee
where $H$ denotes the classical Hamiltonian of the system.

It should be noted that constructing the proper corrections for the canonical average up to a certain order in powers of $\hbar$ requires not only the expansion of the density matrix but also the expansion of the partition function $Z = \mathrm{Tr} \left[ \exp(-\beta \hat{\mathscr{H}}) \right]$, which appears in the denominator of Eq. \eqref{eq:rho_canonical} \cite{nienhuis1970theory}. In what follows, we focus on obtaining the low-order spectral moments, which are sufficient for describing a gas at relatively modest density. As a result, we can assume that 
\be
  \mathrm{Tr}\left[\exp(-\beta\hat{\mathscr{H}})\right] \approx \frac{1}{(2\pi\hbar)^s}\int\int\dd\mathbf{q}\dd\mathbf{p} \exp(-\beta H) \sim V,
\ee
where $V$ is the gas volume.

Non-commutativity of $\hat{q}$ and $\hat{p}$ operators implies that the choice of the distribution function, as given by Eq. \eqref{eq:Pw_wigner}, is not unique \cite{lee1995theory}. This ambiguity is closely related to the choice of the operator ordering, as discussed in Ref. \cite{hillery1984distribution}. The function $A_w$ in Eq. \eqref{eq:A_mean_value} can be represented as a Fourier integral of the form:
\be
  A_w(\mq,\mp) = \iint \alpha(\sigma, \mu)e^{i(\mu\mq+\sigma\mp)}\dd \sigma \dd \mu.
\ee
According to Weyl's rule \cite{weyl1931}, the exponential of the classical variables $e^{i(\mu\mq + \sigma\mp)}$ is associated with the operator $e^{i(\mu\hat{\mq}+\sigma\hat{\mp})}$. This association leads to $P_w$ of the form given by Eq. \eqref{eq:Pw_wigner} in the mean value \eqref{eq:A_mean_value}. Alternatively, another distribution function proposed by Kirkwood \cite{kirkwood1933quantum} can be introduced by adopting a different association rule:
% Provided the exponential of the classical variables $e^{i(\mu\mq+\sigma\mp)}$ is associated with the operator $e^{i(\mu\hat{\mq}+\sigma\hat{\mp})}$ following the Weyl's rule \cite{weyl1931}, the function $P_w$ in Eq. (\ref{eq:A_mean_value}) reduces to the Wigner functon (\ref{eq:Pw_wigner}). Alternatively, another famous distribution function proposed by Kirkwood \cite{kirkwood1933quantum, moyal1949quantum} can be introduced, which arises from a different association rule:
\be
  e^{i(\mu{\mq}+\sigma{\mp})} \longleftrightarrow e^{\sigma\hat{\mp}}e^{i\mu\hat{\mq}}
\ee

Kirkwood's approach can also be applied to derive quantum corrections to classical canonical distributions by means of a power series expansion in Planck's constant \cite{kirkwood1933quantum,hill1956}. Note that the Wigner and Kirkwood expansions become equivalent when integrated over momenta, provided that the phase-space average is taken for a function that depends solely on coordinates.

In order to derive the quantum corrections for the zeroth and first spectral moments, we begin with the Eq. \eqref{eq:Mn_commutator}, which can be rewritten as:
\be
  \mathcal{M}_n = B_n \mathrm{Tr} \big\{  \hat{\rho} \hat{\mathscr{F}}_n \big\},
  \label{eq:F-operator}
\ee
where the operator $\hat{\mathscr{F}}_n$ is defined as:
\setlength\belowdisplayskip{14pt}
\begin{gather}
  \hat{\mathscr{F}}_n = \boldsymbol{\mu}\cdot \smash[b]{ \underbrace{[\mathscr{H},\ldots[\mathscr{H},}_\text{$n$ times}\:\boldsymbol{\mu}]]},
\end{gather}
\setlength\belowdisplayskip{8pt}
and the coefficient $B_n$ is given by:
\begin{gather}
  B_n = V\frac{8\pi^3N_L^2}{3(2\pi\hbar c)^{n+1}}\frac{1}{4\pi\varepsilon_0}.
\end{gather}

By applying the Wigner transform and utilizing the Eq. (\ref{eq:Wigner_AB_trace}), we obtain the following expression for the $n$-th spectral moment:
\be
  \label{eq:Mn_wigner}
  \mathcal{M}_{n}^{w} = \frac{B_n}{(2 \pi \hbar)^s Z} \int\dd \mathbf{q}\int \dd \mathbf{p}\: \Omega_w \lb \hat{\mathscr{F}}_n \rb_w.
\ee
For the zeroth spectral moment, the operator $\hat{\mathscr{F}}_0$ depends only on coordinates, and its Wigner equivalent is simply
\be
  \left(\hat{\mathscr{F}}_0\right)_w = \left(\mu^2(R)  \right)_w = \mu^2(R).
\ee
The first spectral moment requires some derivation, which can be accomplished with the aid of Eq. \eqref{eq:wigner_multiplication}:
\be
\label{eq:F1_expression}
  \left(\hat{\mathscr{F}}_1\right)_w = \left(\boldsymbol{\mu}\cdot[\mathscr{H},\boldsymbol{\mu}]  \right)_w = \frac{\hbar^2}{2m}\sum_{i,j={x,y,z}}\left(\frac{\pp \mu_j}{\pp q_i}   \right)^2.
\ee
For a diatomic system in the absence of external fields, the potential $U$ depends only on the radial coordinate $R$. By integrating Eq.\eqref{eq:Mn_wigner} over the momenta and the orientation of the interatomic axis, we obtain the following expressions:
 
\begin{widetext}
\begin{align}
   \mathcal{M}_0^w &= (2\pi m k_\text{B} T)^{3/2} \frac{4 \pi B_0}{(2 \pi \hbar)^s Z} \int\limits_0^\infty \: f_w(R) \mu^2(R) R^2\dd R, \label{eq:zeroth_wigner}\\
  \mathcal{M}_1^w &= (2\pi m k_\text{B} T)^{3/2} \frac{4 \pi B_1}{(2 \pi \hbar)^s Z} \frac{\hbar^2}{2m} \int\limits_0^\infty \: f_w(R) \left\{\left(\frac{\dd \mu}{\dd R}\right)^2+\frac{2}{R^2}(\mu(R))^2\right\} R^2\dd R,\label{eq:first_wigner}
\end{align}
\end{widetext}
where $f_w(R)$ is defined as
\be
\begin{aligned}
  f_w(R) &= \int \Omega_w(\mathbf{q}, \mathbf{p}) \, \dd\mp \, \dd \theta \, \dd\phi = \\ 
  &\Big[ 1 + \hbar^2 \xi_2(R) + \hbar^4\xi_4(R) + \ldots \Big] e^{-\beta U(R)}.
\end{aligned}
\ee
In order to derive the expressions for the quantum corrections to the density operator, we represent $\Omega_w$ as given in the Eq. \eqref{eq:Omega_expansion} and expand the cosine in Eq. \eqref{eq:Omega_w_cos} as a power series in $\hbar$. By equating the coefficients of corresponding powers of $\hbar$ on both sides of the equation, we obtain a set of equations that can be solved for the correction terms:
\begin{gather}
\begin{aligned}
  \hbar^0: &\;-H\exp(-\beta H) = -H\exp(-\beta H), \\
  \hbar^2:&\; -H\chi_2\exp(-\beta H)+\frac{\partial \chi_2}{\partial \beta}\exp(-\beta H) = \\
  &-H\chi_2\exp(-\beta H)+H\frac{\Lambda^2}{8}\exp(-\beta H). \\
  % \hbar^4:& \qquad \ldots
\end{aligned}
\end{gather}
In deriving these equations, we employed the fact that $H_w(\mq, \mp) = H(\mq, \mp)$. As can be seen, the zeroth-order term simply yields a trivial identity. The expression for $\chi_2$ is derived by solving the differential equation that arises at second order. 

With the aid of computer algebra system Maple \cite{maple}, we have derived closed expressions for $\chi_2$, $\chi_4$, $\chi_6$, $\chi_8$ and then $\xi_2$, $\xi_4$, $\xi_6$ and $\xi_8$. The corrections for spectral moments are effected by two types of corrections: static corrections, arising from the density operator, and dynamic corrections, arising from the commutators in Eq. \eqref{eq:Mn_commutator}.

The expressions for $\xi_n$ and $\chi_n$ up to the order of $\hbar^6$ were derived previously in Refs. \cite{haberlandt1974quantenstatistik, Barocchi1987}. Our derived expressions are in complete agreement with those presented in Ref. \cite{Barocchi1987}, but differ from those in Ref. \cite{haberlandt1974quantenstatistik} at the $\hbar^6$ order, suggesting a possible typographical error in the latter reference. To provide a reliable reference for future research, we have included our derived corrections up to $\hbar^8$ in Cartesian coordinates in Supplementary material, encoded as \texttt{sympy} strings. We believe that providing these corrections in a machine-readable format is more advantageous compared to traditional mathematical formulas, as it eliminates the tedious and error-prone process of translating equations into code, thereby reducing the risk of implementation errors and facilitating the incorporation of high-order quantum corrections into calculations.
% We can only speculate that erroneous value of the $\hbar^6$ order correction reported in \cite{haberlandt1974quantenstatistik} might be a result of a typographical error. 

\subsection{Path integral formulation}
\label{subsec:path-integral}

The path integral method has forged itself as a powerful tool for the investigation of many-body quantum systems. This approach has been successfully applied in studies of off-lattice condensed-phase systems, such as doped clusters \cite{Kwon2000} and molecular systems within otherwise confined geometries \cite{Ramirez1995}.    
While considering the gas phase effects, the path integral methods were successfully employed in calculations of the virial coefficients for the low-temperature monoatomic gases, for which quantum effects become pronounced \cite{Fosdick1966}. Later on such calculations were extended, taking into account rotational degrees of freedom, e.g., for H$_2-$H$_2$ \cite{Patkowski2008}. 

Conventionally, the ensemble expectation value of a quantum-mechanical operator $\hat{O}$ is expressible as follows:
\begin{gather}
    \langle \hat{O} \rangle = \text{Tr} \lc \hat{\Omega} \, \hat{O} \rc / \text{Tr} \lc \hat{\Omega} \rc.
\end{gather}
In the simplest cases, when an observable $\hat{O}$ depends solely on atomic positions and a system of nuclei can be regarded as distinguishable, the canonical average $\avg{\hat{O}}$ can be represented within Feynman's path integral formalism as \cite{Tuckerman2002} 
\begin{widetext}
\begin{gather}
\begin{aligned}
    \avg{\hat{O}} = \frac{1}{Z} \lim_{P \rightarrow \infty} \lb \frac{m P}{2 \pi \beta \hbar^2} \rb^{3P/2} \int \diff{\mf{x}}_1 \dots \diff{\mf{x}}_P \,   o_P(\mf{x}_1, \dots, \mf{x}_P) \exp \lc -\sum_{i = 1}^{P} \left[ \frac{m P}{2 \beta \hbar^2} (\mf{x}_{i + 1} - \mf{x}_{i})^2 + \frac{\beta}{P} U(\mf{x}_i) \right] \rc_{\mf{x}_{P + 1} = \mf{x}_1}.
\end{aligned}
  \label{eq:feynman-average}
\end{gather}
\end{widetext}
Note that the quantity 
\begin{gather}
    o_P(\mf{x}_1, \dots, \mf{x}_P) = \frac{1}{P} \sum_{i = 1}^{P} o(\mf{x}_i)
\end{gather}
is an estimator for the operator $\hat{O}$, which means that for large $P$ values its average across the probability distribution
%\begin{widetext}
\begin{align}
    \rho_P(\mf{x}_1, \dots, &\mf{x}_P) = \frac{1}{Z_P} \lb \frac{m P}{2 \pi \beta \hbar^2} \rb^{3P/2} \times \notag \\
    &\exp \lc -\sum_{i = 1}^{P} \left[ \frac{m P}{2 \beta \hbar^2} (\mf{x}_{i + 1} - \mf{x}_{i})^2 + \frac{\beta}{P} U(\mf{x}_i) \right] \rc
    \label{eq:feynman-density}
\end{align}
%\end{widetext}
%
which is analogous to a Boltzmann density, yields canonical expectation value of an operator $\hat{O}$.

Importantly, in our considered case of a heteroatom pair, the spin statistics can be ignored.
The Feynman path integral establishes an isomorphism between the quantum-mechanical expectation value and the classical average over the configurational space of a cyclic polymer made up of $P$ atoms, each subject to potential $U$ and harmonic attractive force with spring constant $m P / (2 \beta \hbar^2)$ exerted by its nearest neighbors. This collection of atoms is commonly referred to as a \textit{necklace}, with the individual particles denoted as \textit{beads}. 
\iffalse
The latter formula employs a primitive approximation for expanding a product of kinetic and potential terms in the density matrix is utilized. 
\fi
Here, we will describe an approach for obtaining the path integral estimates for zeroth and first spectral moments for a diatomic system with a Hamiltonian shown by Eq. \eqref{eq:radial-schroedinger}. The average \eqref{eq:feynman-average} is evaluated in Cartesian coordinates for a necklace of atoms, each with a reduced mass $m$ and subject to potential $U$. Note that particle positions $\mf{x}$ are described by three-dimensional vectors.

\begin{figure}[htbp]
    \centering
    \includegraphics[width=0.6\linewidth]{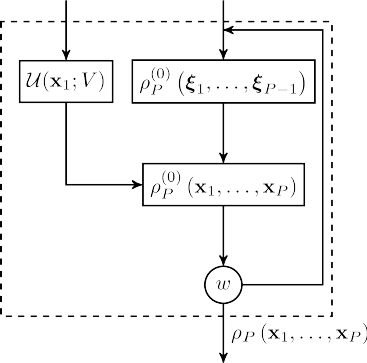}
    \caption{Schematic representation of the Monte Carlo algorithm producing unweighted samples according to distribution function $\rho_P$.}
    \label{fig:monte-carlo-scheme}
\end{figure}

The estimators for operators $\hat{\mathscr{F}}_0$ and $\hat{\mathscr{F}}_1$ which are directly related to the zeroth and first spectral moments by Eq. \eqref{eq:F-operator} are given by 
\begin{gather}
\begin{aligned}
    \mathscr{F}_{0, P}(\mf{x}_1, \dots, \mf{x}_P) &= \frac{1}{P} \sum_{i = 1}^{P} \boldsymbol{\mu}^T(\mf{x}_i) \boldsymbol{\mu}(\mf{x}_i), \\
    \mathscr{F}_{1, P}(\mf{x}_1, \dots, \mf{x}_P) &= \frac{1}{P} \frac{\hbar^2}{2 m} \sum_{i = 1}^{P} \sum_{j, k} \lb \nabla_j \boldsymbol{\mu}_k \rb^\top \lb \nabla_j \boldsymbol{\mu}_k \rb.
\end{aligned}
\end{gather}
The interparticle potential $U$ for our system in question is extremely weak, and the spectral effects are largely dominated by free-free transitions even at relatively low temperatures. Because of this, the procedure of sampling the beads coordinates has to be specifically designed to be able to deal with unbound states. We have developed the sampling procedure from the density \eqref{eq:feynman-density} using a multistep algorithm represented schematically in Figure \ref{fig:monte-carlo-scheme}. First, one of the beads is chosen as a reference, be it, for example, $\mf{x}_1$, and a set of difference coordinates is introduced relative to those of the reference bead 
\begin{gather}
    \boldsymbol{\xi}_i = \mf{x}_{i + 1} - \mf{x}_1, \quad i = 1, \dots P - 1.
\end{gather}
Next, a proposal distribution $\rho_P^{(0)}$ is derived from the target distribution $\rho_P$ by assuming the potential $U$ is zero. The unnormalized distribution function is expressed in terms of relative coordinates as follows:
\begin{widetext}
\begin{gather}
    \rho_P^{(0)}(\boldsymbol{\xi}_1, \dots, \boldsymbol{\xi}_{P - 1}) \propto \exp \lc -\frac{m P}{2 \beta \hbar^2} \lsq \boldsymbol{\xi}_1^\top \boldsymbol{\xi}_1 + \boldsymbol{\xi}_{P - 1}^\top \boldsymbol{\xi}_{P - 1} + \sum_{i = 1}^{P - 2} \lb \boldsymbol{\xi}_{i + 1} - \boldsymbol{\xi}_i \rb^\top \lb \boldsymbol{\xi}_{i + 1} - \boldsymbol{\xi}_i \rb \rsq \rc.   
    \label{eq:difference-distribution}
\end{gather}
\end{widetext}
The boxes in the sketch (Fig.\ref{fig:monte-carlo-scheme}) denote algorithms that can be initially regarded as black boxes, generating unweighted samples of relevant variables in accordance with the distribution functions specified within each of the boxes. The set of relative coordinates sampled from $\rho_P^{(0)}(\boldsymbol{\xi}_1, \dots, \boldsymbol{\xi}_{P - 1})$ are combined with the variable $\mf{x}_1$ to yield a sample of necklace coordinates that follow $\rho_P^{(0)}(\mf{x}_1, \dots, \mf{x}_P)$. Subsequently, a rejection step, schematically represented by a circle, is applied to necklace samples. For each sample, a weight $w$ is calculated, and the sample is accepted (depicted by the arrow coming outside of the dashed box) if $u W < w$, where $u$ is uniform $u \sim \mathcal{U}\lsq 0, 1 \rsq$ and $W$ is a maximum weight, and rejected otherwise (depicted by the return arrow). The weight $w$ is calculated as the ratio of target density to the proposal density:
\begin{gather}
    w = \frac{\rho_P(\mf{x}_1, \dots \mf{x}_P)}{\rho_P^{(0)}(\mf{x}_1, \dots \mf{x}_P)} = \exp \lc -\frac{1}{P} \sum_{i = 1}^{P} \beta U(\mf{x}_i) \rc,
\end{gather}
with a maximum weight given by $W = \exp (- U_\text{min} / k_\text{B} T)$, where $U_\text{min}$ is the global minimum of the interparticle potential.

The introduction of relative variables is motivated by their roughly Gaussian marginal distributions at temperatures between 50 and 300 K, which facilitates robust sampling via Markov chain techniques. The beads coordinates then constructed from these difference variables turn out to have marginal distributions that only slightly differ from the uniform. This demonstrates how challenging direct sampling is using conventional Monte Carlo techniques. 

\section{Discussion}
\label{sec:discussion}

We performed extensive quantum-mechanical, quantum-statistical, and trajectory-based calculations employing the interparticle potential and induced dipole functions for He$-$Ar reported by \citet{Cacheiro2004}. Specifically, the fitting function for the potential was used as proposed by \citet{Korona1997} with parameters reported in Table 2 of Ref. \cite{Cacheiro2004}. Analytical function was fitted to reproduce the induced dipole values reported in Table 4 of Ref. \cite{Cacheiro2004}, and the resulting parameters are summarized in Table \ref{tab:dipole}. The routines in \texttt{C} implementing the interparticle potential and induced dipole are provided in Supplementary material.

\begin{table}
    \centering
    \caption{Fitting parameters of induced dipole for He$-$Ar, represented in the form $\mu(R) = A \exp(-\alpha R - \beta R^2) - c_7/R^7$. Atomic units are used for the interparticle distance and resulting dipoles.}
    \begin{tabular}{p{3cm}p{3cm}}
      \toprule
        & He$-$Ar \\
      \midrule
      $A$, a.u. &  1.6046920 \\
      $\alpha$, $a_0^{-1}$ &  0.40100613 \\
      $\beta$, $a_0^{-2}$ & 0.10292726 \\
      $c_7$, $a_0^7$ & 148.55391 \\
      \bottomrule
    \end{tabular}
    \label{tab:dipole}
\end{table}

\begin{widetext}
    
\begin{figure}[!htbp]
    \centering
    \includegraphics[width=0.49\linewidth]{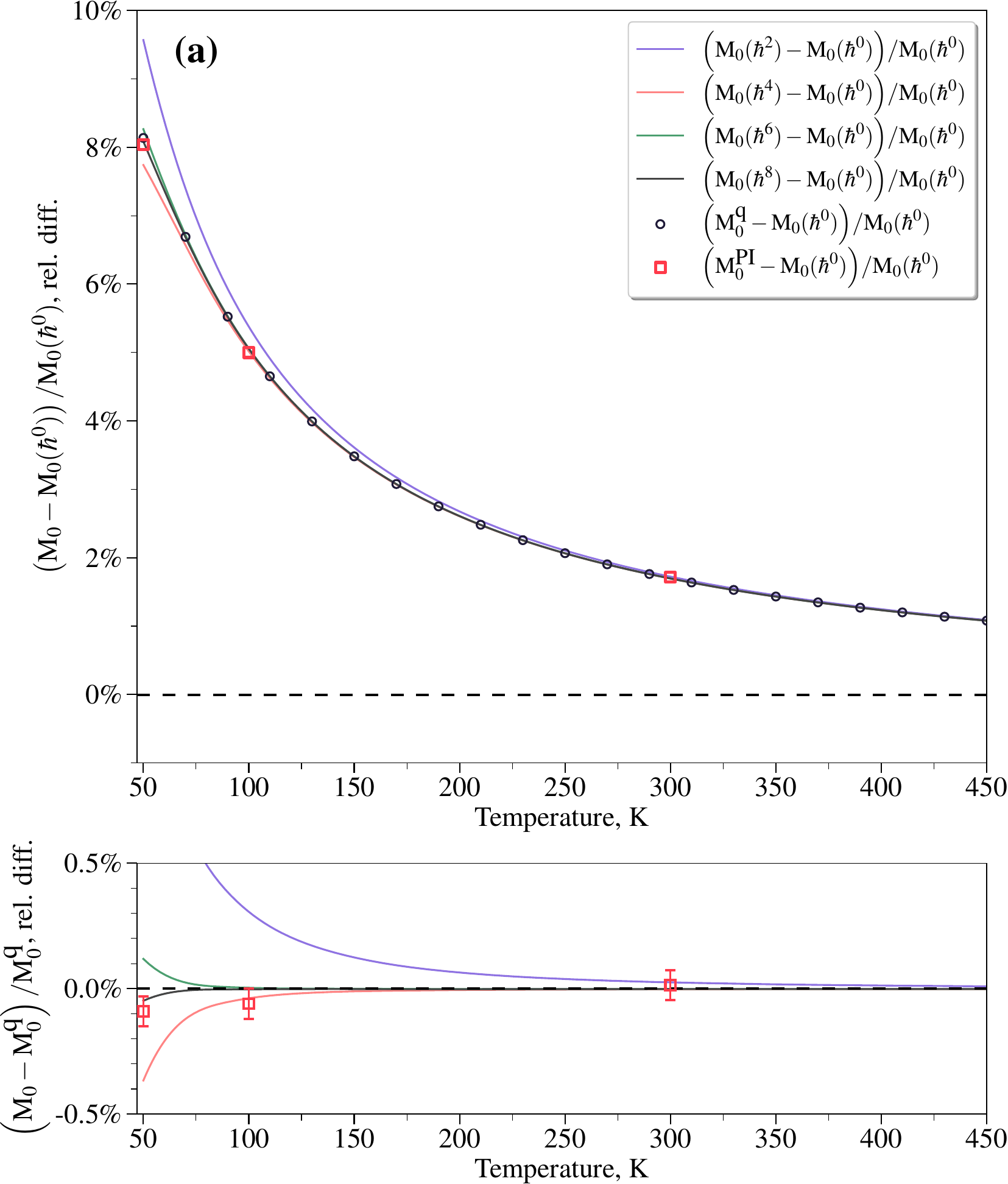}
    \includegraphics[width=0.49\linewidth]{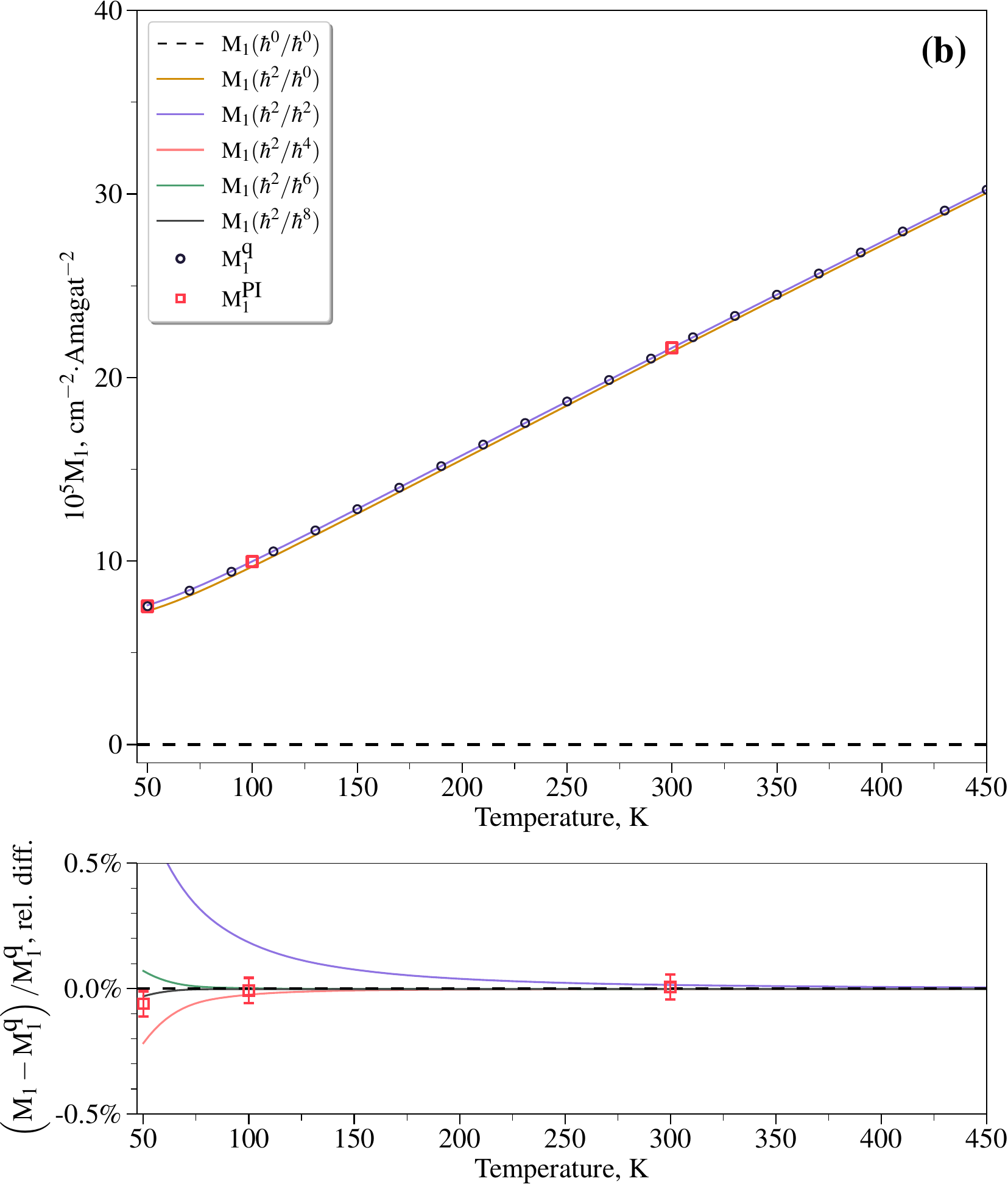}
    \caption{Temperature variation of the relative deviation of the zeroth spectral moment from the classical value (a, top) and from the quantum-mechanical value (a, bottom), and the absolute value of the first spectral moment (b, top) and its relative deviation from the quantum-mechanical value (b, bottom). The colored solid lines represent the Wigner expansion results with corrections to the density operator up to $\hbar^2$ (blue), $\hbar^4$ (red), $\hbar^6$ (green), and $\hbar^8$ (black). For the first spectral moment, the notation is split to separately denote the orders of the dynamic correction and the static correction to the density operator. The yellow line corresponds to the first spectral moment computed with dynamic correction but without any static corrections. The time-independent quantum-mechanical formalism values are shown as open circles, while path integral approach values with the maximum available $P$ are denoted by red squares. Note that M$_1$ is zero when considered within classical formalism. (For interpretation of the references to color in this legend, the reader is referred to the web version of the article.)}
    \label{fig:moments}
\end{figure}
\end{widetext}

Throughout all calculations in this work, the reduced mass of the He-Ar pair was taken as $m = 6632.039\: m_e$, where $m_e$ denotes the electron mass. The results of our conducted spectral moments calculations are summarized in Table \ref{tab:spectral-moments}. A graphical representation of the deviations between the computed estimates and corresponding classical values for the zeroth and first spectral moments  is shown in Figure \ref{fig:moments}.

\begin{table*}[!ht]
  \centering
  \caption{Zeroth and first spectral moments of He-Ar CIA band at 50, 100 and 300 K obtained using various approaches}
  \begin{threeparttable}
  \begin{tabular}{cl@{\hspace{0.75cm}}ccc@{\hspace{0.75cm}}ccc}
    
    \toprule
    & & \multicolumn{3}{c}{M$_0$, cm$^{-1}\cdot$Amagat$^{-2}$} & \multicolumn{3}{c}{M$_1$, cm$^{-2}\cdot$Amagat$^{-2}$} \\
     & & 50 K & 100 K & 300 K & 50 K & 100 K & 300 K \\
    \midrule
    \multirow{2}{*}{Quantum} & a. sum formula & 1.5888e-06 & 2.2279e-06 & 5.4551e-06 & 7.5374e-05 & 9.9651e-05 & 2.1613e-04 \\
                             & b. spectrum & 1.5874e-06 & -- & 5.4533e-06 & 7.5224e-05 & -- & 2.1500e-04 \\
    \midrule
    \multirow{2}{*}{Classical} & a. phase-space & 1.4693e-06 & 2.1206e-06 & 5.3639e-06 & 0 & 0 & 0 \\
                               & b. spectrum & 1.4287e-06 & -- & 5.3492e-06 & 0 & 0 & 0 \\
    \midrule
    \multirow{2}{*}{Desymmetrization} & a. D3 & --$^{a}$ & -- & 5.6133e-06 & --$^{a}$ & -- & 2.2860e-04 \\
                                      & b. D4a & 1.9104e-06 & -- & 5.5996e-06 & 8.2250e-05 & -- & 2.1628e-04 \\
    \midrule
    \multirow{5}{*}{Path integral} & a. $P = 2$ & 1.5325e-06 & 2.1913e-06 & 5.4402e-06 & 7.4151e-05 & 9.8816e-05 & 2.1573e-04 \\
                          & b. $P = 4$ & 1.5678e-06 & 2.2178e-06 & 5.4568e-06 & 7.4819e-05 & 9.9393e-05 & 2.1622e-04 \\
                          & c. $P = 8$ & 1.5822e-06 & 2.2239e-06 & 5.4559e-06 & 7.5172e-05 & 9.9563e-05 & 2.1614e-04 \\
                          & d. $P = 16$ & 1.5874e-06 & 2.2266e-06 & -- & 7.5230e-05 & 9.9648e-05 & -- \\
                          & e. $P = 32$ & -- & -- & -- & 7.5329e-05 & -- & -- \\
                          % & e. $P = \infty$ &  1.5874e-06 & 2.2276e-06 & 5.4552e-06 & 7.5176e-05 & 9.9622e-05 & 2.1638e-04 \\

    \midrule
    \multirow{4}{*}{Wigner expansion} & a. $\hbar^2$ & 1.6099e-06 & 2.2347e-06 & 5.4565e-06 & 7.2775e-05 & 9.6996e-05 & 2.1408e-04 \\
                            & b. $\hbar^4$ & 1.5830e-06 & 2.2270e-06 & 5.4550e-06 & 7.5920e-05 & 9.9835e-05 & 2.1616e-04 \\
                            & c. $\hbar^6$ & 1.5907e-06 & 2.2279e-06 & 5.4551e-06 & 7.5211e-05 & 9.9628e-05 & 2.1612e-04 \\
                            & d. $\hbar^8$ & 1.5881e-06 & 2.2278e-06 & 5.4551e-06 & 7.5428e-05 & 9.9653e-05 & 2.1612e-04 \\
                            & e. $\hbar^{10}$& -- & -- & -- & 7.5351e-05 & 9.9649e-05 &2.1612e-04\\
    
    \bottomrule
  \end{tabular}
  \begin{tablenotes}
    \item[a] The D3 desymmetrization procedure applied to the trajectory-based spectrum at 50 K results in a profile exhibiting unphysical behavior (see Figure \ref{fig:spectra} and discussion in Section \ref{sec:discussion}), and therefore, spectral moments were not calculated in this case. 
  \end{tablenotes}
  \end{threeparttable}
\end{table*}
{\makeatletter\def\@currentlabel{\thetable}\label{tab:spectral-moments}}

\begin{figure}[!htbp]
    \centering
    \includegraphics[width=\linewidth]{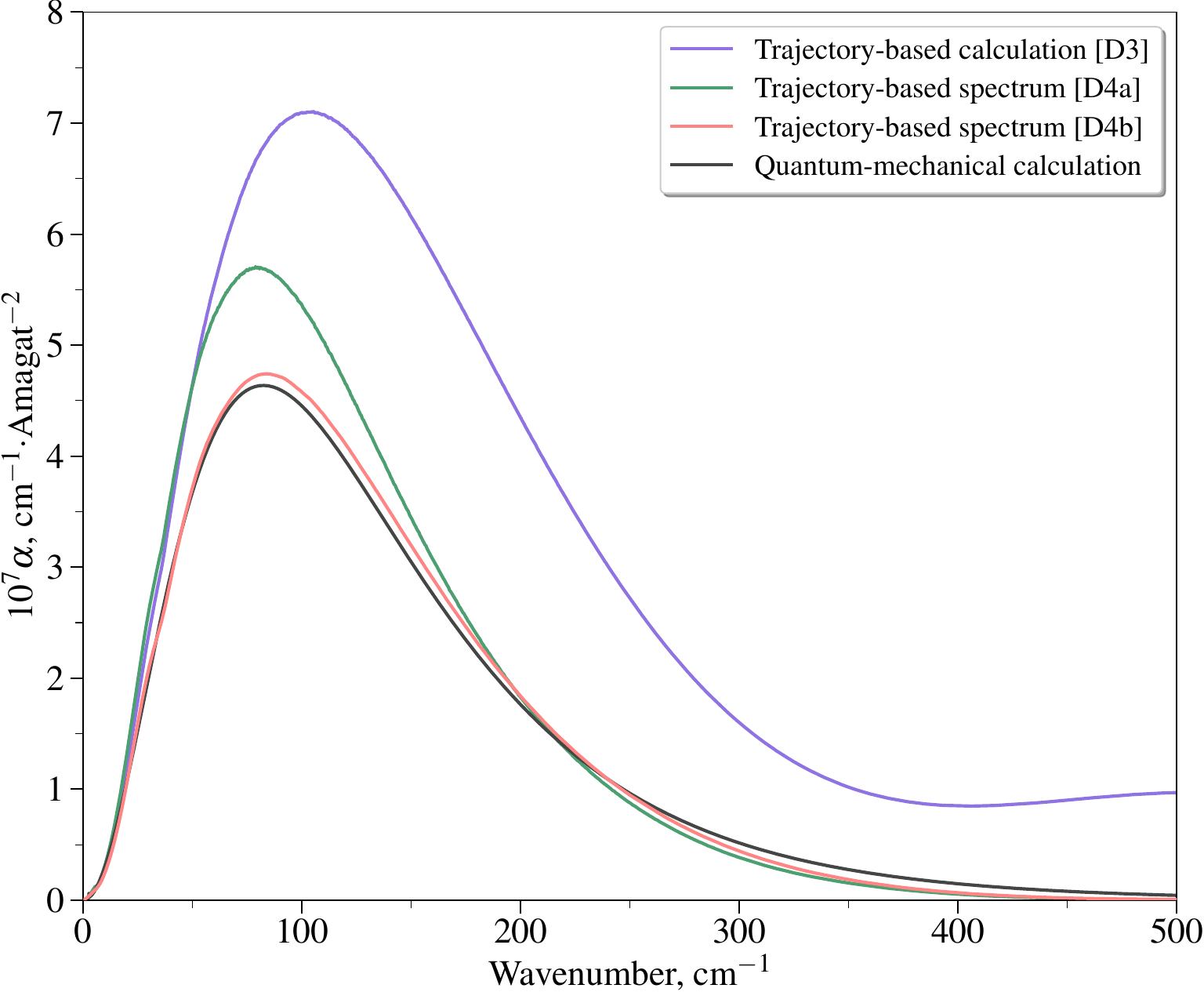}
    \includegraphics[width=\linewidth]{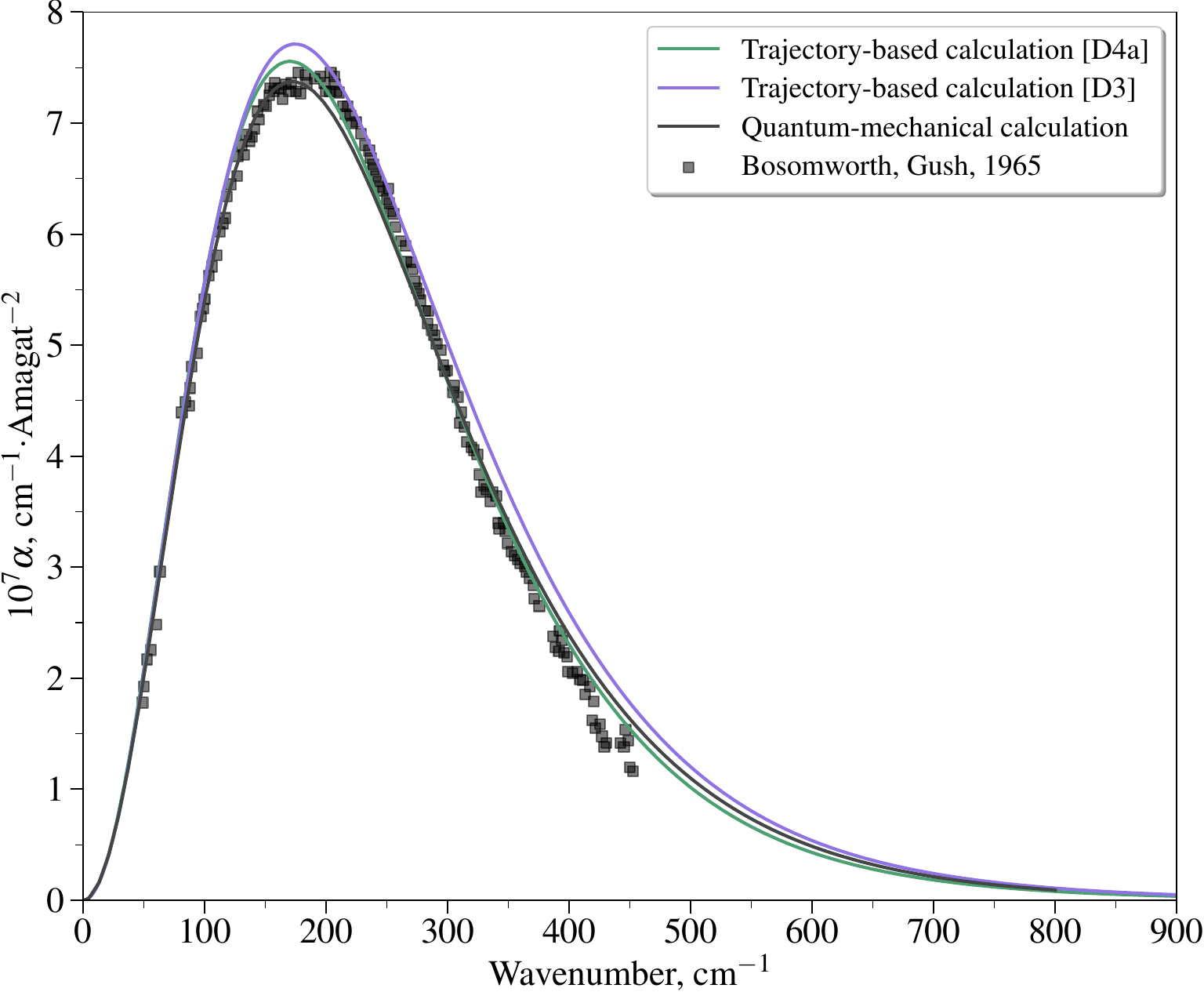}
    \caption{He-Ar CIA spectra at 50 K (top panel) and 300 K (bottom panel). The black line shows the result of quantum-mechanical calculation. The results of trajectory-based calculation using Sch\"ofield's, Frommhold's and our extended D4b desymmetrization are represented using blue, green and red lines, respectively. Black squares show the measured data of \citet{Bosomworth1965}. (For interpretation of the references to color in this legend, the reader is referred to the web version of the article.)}
    \label{fig:spectra}
\end{figure}

The path integral estimates of the zeroth and first spectral moments were computed according to the procedure outlined in Section \ref{subsec:path-integral}. Sampling from the difference distribution, as defined by Eq. \eqref{eq:difference-distribution}, was performed via the Markov chain Monte Carlo approach, utilizing Metropolis-Hastings algorithm \cite{Metropolis1953}. To achieve a statistical uncertainty of obtained spectral moments estimates of less than 0.05\% for each value of $P$, the sample sizes ranging from 1 to 30 billion were utilized. The convergence to the limiting value with $P$ is monotonic, and the number of beads required to reach an estimate with a given precision is, to a first order, inversely proportional to the temperature. The path integral method yields satisfactory estimates of the spectral moments already at lower values of $P$. However, by pushing the method to achieve an estimate with precision of 0.05\%, we demonstrate its capabilities, notwithstanding that such an extensive sample size (up to tens of billions) may not be reasonable in typical calculations.

The classical values and Wigner corrections for the zeroth and first spectral moments were calculated using Eqs. \eqref{eq:zeroth_wigner} and \eqref{eq:first_wigner}. The integrals were computed using the \texttt{VEGAS} variant of the adaptive Monte-Carlo algorithm \cite{Lepage1978,VEGAS}. The statistical uncertainty of about 0.005\% was achieved. Note that dynamic correction for the first spectral moment, expressed by Eq. \eqref{eq:F1_expression}, is already of order $\hbar^2$. Consequently, incorporating corrections to the density operator up to $\hbar^8$ results in overall corrections extending up to $\hbar^{10}$.

Quantum-mechanical calculations for the zeroth and first spectral moments were carried out using the Eqs. \eqref{eq:zeroth_g0} and \eqref{eq:first_g0}, respectively. Accordingly, quantum mechanical spectral functions were calculated using Eqs. (\ref{eq:J_ff_quan})-(\ref{eq:J_fb_quan}). To obtain the bound states wavefunctions, we employed the matrix Numerov algorithm \cite{pillai2012matrix}. The chosen potential supports bound states up to an end-over-end angular momentum of $\ell = 4$. The wavefunctions for the free states were propagated to 150 $a_0$ where they were matched to the form
\be
  \psi_i(R;E_i,\ell_i) \sim kR \Big[ j_{\ell_i}(kR)\cos\delta_{\ell_i}-n_{\ell_i}(kR)\sin\delta_{\ell_i} \Big],
\ee
where $k^2 = 2mE/\hbar^2$, $j_{\ell}$ and $n_{\ell}$ are the spherical Bessel functions of the first and second kind, respectively. In propagating the free states wavefunctions, a radial grid consisting of 16,000 points was employed. 
The integral over energies in Eqs. \eqref{eq:J_ff_quan} and \eqref{eq:g0_quantum} were evaluated using Simpson's rule with an upper limit of 7,000 cm$^{-1}$. To achieve high accuracy, the energy range was divided into two segments. The first segment, spanning from 0 to 25 cm$^{-1}$, was discretized using 4,000 points  for the calculation of spectral moments and 1,500 points for the calculation of the spectral function. The second segment, ranging from 25 cm$^{-1}$ to 7,000 cm$^{-1}$, was represented with 900 points.

The terms of Wigner expansion exhibit alternating signs, analogous to corrections to the second virial coefficient \cite{mason1969virial}. Notably, as demonstrated by the lower panels on Figure \ref{fig:moments}, the quantum-mechanical value falls within the range between the cumulative values up to $\hbar^{2n}$ and $\hbar^{2n + 2}$.
Within the temperature range of 50$-$500 K, the expansion series converges rapidly. Interestingly, the corrections up to $\hbar^6$ suffice for a practical estimate, even at the lowest temperatures. 
For the zeroth spectral moment, the partial sum of the Wigner expansion series up to $\hbar^8$ agrees with both the quantum-mechanical value and the path integral estimate to within 0.05\%, at reference temperatures of 50, 100, and 300 K.
Furthermore, the first spectral moment, which has a classical value of zero, is already satisfactorily approximated by the first correction term, accurate up to $\hbar^2$. However, it should be noted that this initial correction term does not appear on the lower panel of Figure \ref{fig:moments} due to its deviation from the quantum-mechanical values exceeding 0.5\%. For the first spectral moment, the partial sum of the Wigner expansion series up to $\hbar^{10}$ also agrees with the quantum-mechanical value and path integral estimate to within 0.05\%. 

\subsection{Trajectory-based spectral simulation}
\label{subsec:trajectory}

In this section we suggest an improvement of the desymmetrization procedure aimed at a significant increase in the accuracy of the spectral profile simulation, especially at low temperatures. This improvement is based on the use of the zeroth and the first spectral moments properly corrected as described in the precedent sections. 
The trajectory-based calculations of the correlation functions were conducted at 50 and 300 K, having propagated 300 or 500 million trajectories, respectively, for unbound states and up to 1 million trajectories for true bound states. The correlation functions along individual trajectories were sampled at fixed time intervals of 4.8 fs. Zeroth spectral moments derived from the resulting profiles using Eqs. \eqref{eq:mom_q_alpha} are shown in Table \ref{tab:spectral-moments}. Figure \ref{fig:spectra} shows the spectral profiles obtained after applying desymmetrization procedures of Sch\"ofield (Eq. \eqref{eq:des-schofield}), Frommhold (Eq. \eqref{eq:des-frommhold}), and its extended version, which will be described below. At room temperature, an impressive agreement can be seen between the desymmetrized trajectory-based profiles, the quantum-mechanically calculated profile, and laboratory observations. However, at 50 K, the limitations of the desymmetrization procedures become apparent. It is seen that both Sch\"ofield's and Frommhold's procedures notably overestimate the quantum-mechanical profile. Moreover, the use of the Sch\"ofield's procedure results in an unphysical artifact in the far wing of the band. 

Armed with the knowledge of the low-order spectral moments properly corrected for quantum effects as described above, we can suggest a new modification of the desymmetrization procedure. We propose the following extension of Egelstaff's procedure, which is referred to hereafter as D4b: 

\begin{gather}
\begin{aligned}
  J_\text{D4b}(\nu) = &d_0 \exp \lb \frac{h c \nu}{2 k_\text{B} T} \rb \times \\ 
  &\int\limits_{-\infty}^{+\infty} C_\text{cl} \lb \sqrt{t^2 + d_1\lb \beta \hbar / 2 \rb^2} \rb e^{-2 \pi i c \nu t} \diff{t}, 
\end{aligned}
\end{gather}
where $d_0$, $d_1$ are adjustable parameters. Note that setting $d_0 = C_\text{cl}(0)/C_\text{cl}(\beta h c / \sqrt{2})$ and $d_1 = 1$ reduces the D4b procedure to D4a (see Eq. \eqref{eq:des-frommhold}).
Starting with the classical correlation function corresponding to the totality of unbound and bound states at 50 K, the parameters $d_0$ and $d_1$ were fitted using the Broyden–Fletcher–Goldfarb–Shanno algorithm in such a way that the zeroth and first spectral moments of the desymmetrized spectral function have to match the values determined with the most performant quantum-statistical approaches. It can be easily shown that the resulting $J_\text{D4b}$ satisfies quantum detailed balance. As shown in Figure \ref{fig:spectra}, this extended D4b desymmetrization permits obtaining the spectral profile in excellent agreement with the quantum-mechanical profile at 50 K, which is incomparably much more accurate than any other desymmetrized profiles. 

\section{Conclusions}
\label{sec:conclusions}

A comprehensive study of quantum corrections to the zeroth and first spectral moments for the CIA band was conducted. Thorough reconsideration of the analytical expressions for the $\hbar$-expansion of the canonical density matrix guided by a computer algebra system revealed possible imperfections in some results reported previously by \citet{Haberlandt1974}. The $\hbar^8$-order correction term is derived in our work for the first time. The $\hbar$-expansion converges rapidly for the zeroth and first spectral moments of the He-Ar CIA band over the temperature range 50$-$500 K. Moreover, practical estimates can be easily obtained even at the lowermost temperatures. The results obtained using Wigner expansion are corroborated by both extensive quantum-mechanical calculations and path integral estimates. These findings can be regarded as an initial step toward the possibility of simulating CIA profiles employing Moyal quantum mechanics formulation \cite{Osborn1995,McQuarrie2006} as an approximation to quantum dynamics. 

Furthermore, as a showcase, an extension of Egelstaff's procedure is proposed to generate spectral profile using quantum-corrected lower-order spectral moments. Starting from a classical profile, determined through trajectory-based calculation, and the quantum values of zeroth and first spectral moment, an adjustable bandshape is defined. The resulting profile, which accurately matches the quantum zeroth and first moments, exhibits an impressive agreement with the results of quantum-mechanical consideration, even at temperatures as low as 50 K.

\begin{acknowledgments}

We were able to carry out trajectory-based calculations and path integral calculations thanks to HPC facilities available at the A.M. Obukhov Institute of Atmospheric Physics, RAS as well as the partial support from the state assignment No. 125020501413-6 to the A. M. Obukhov Institute of Atmospheric Physics, RAS. D.N.C. acknowledges the support by RSCF Grant No. 24-77-00072 for conducting calculations within quantum mechanical and Wigner phase space frameworks. A.A.F and D.N.C. are thankful to A.A. Vigasin and M.E. Gorbunov for insightful discussions.
\end{acknowledgments}

%\section{Supplementary material}

\bibliography{biblio}

\end{document}